\documentclass{kluwer}    

\newdisplay{guess}{Conjecture}
\usepackage{graphicx}

\begin{document}
\begin{article}
\begin{opening}
\title{Model of the W3(OH) environment based on data for both maser and \lq quasi-thermal\rq\ methanol lines}
\runningauthor{A.M. Sobolev et al.}
\runningtitle{Model of W3(OH) environment}

\author{Andrei M. \surname{Sobolev}}
\institute{Ural State University, Russia}
\author{Edmund C. \surname{Sutton}}
\institute{University of Illinois at Urbana-Champaign, USA}
\author{Dinah M. \surname{Cragg}}
\institute{Monash University, Australia}
\author{Peter D. \surname{Godfrey}}
\institute{Monash University, Australia}
\date{April 15, 2004}

\begin{abstract}
   In studies of the environment of massive young stellar objects, recent progress in both observations and theory allows a unified treatment of data for maser and \lq quasi-thermal\rq\ lines. Interferometric maser images provide information on the distribution and kinematics of masing gas on small spatial scales. Observations of multiple masing transitions provide constraints on the physical parameters.

   Interferometric data on \lq quasi-thermal\rq\ molecular lines permits an investigation of the overall distribution and kinematics of the molecular gas in the vicinity of young stellar objects, including those which are deeply embedded. Using multiple transitions of different molecules, one can obtain good constraints on the physical and chemical parameters.
   Combining these data enables the construction of unified models, which take into account spatial scales differing by orders of magnitude.

   Here we present such a combined analysis of the environment around the ultracompact HII region in W3(OH). This includes the structure of the methanol masing region, physical structure of the near vicinity of W3(OH), detection of new masers in the large-scale shock front and embedded sources in the vicinity of the TW young stellar object.
\end{abstract}

\end{opening}

\section{Possibility of combined analysis of data on maser and \lq quasi-thermal\rq\ lines}
Modern studies of the kinematical, physical and chemical status of star forming regions are largely based on the information delivered by emission in molecular lines. Different molecular transitions have different excitation temperatures, as a result of the radiative and collisional population exchange between quantum levels. In the stationary case these take the form of pumping cycles (see, e.g., \opencite{sob94}). Excitation states of molecular transitions are mostly characterized by positive excitation temperatures and, hence, the optical depths of the corresponding lines are mostly positive. Such lines are called \lq quasi-thermal\rq\ and trace the whole extent of the sufficiently dense and warm environment of the massive young stellar object where the corresponding molecule is abundant.

Under specific conditions, in some regions of a molecular cloud the pump cycles can lead to extreme overheating of some quantum transitions and their excitation temperatures might become negative (see, e.g., \opencite{sob94}). In this case the optical depth of the corresponding line becomes negative and emission from material which is more distant from the observer is amplified at the transition frequency instead of being hidden. Such extreme deviation from the local thermodynamic equilibrium (LTE) is called a maser and often leads to extremely bright lines arising from compact areas which can be substantially smaller in size than the region where the maser is formed (see, e.g., \opencite{sob98}).

At present, combined analysis of the data in both types of lines
(i.e., maser and  \lq quasi-thermal\rq\ ) is uncommon. The main
reason is that the spatial resolutions used in the maser research
(rather small fractions of an arcsecond, down to microarcseconds) and
in the \lq quasi-thermal\rq\ line research (arcseconds and tens of
arcseconds) are greatly different.

From the time of its discovery, maser emission has provided very important information on the nature and kinematics of associated objects. Despite this, the structure of the masing regions themselves is rather unclear; these are the regions where maser amplification may develop along particular lines of sight due to the existence of population inversion. For example, current models of maser excitation meet great difficulties in trying to produce extremely bright masers in regions matching the sizes of the tiny maser hot spots which are determined interferometrically. At the same time there are indications that the masing regions  have extensions much greater than these spot sizes.  Indeed, analysis of multi-transitional observations of a masing molecule such as methanol (CH$_3$OH) shows that the sizes of the masing regions can be comparable to those of the associated molecular cores (see, e.g., \opencite{sob93}, \opencite{sly}). There are indications that strong masers also form in rather extended regions. Firstly, VLBI observations show that the strongest H$_2$O and CH$_3$OH masers have a \lq core - extended halo\rq\ structure (see, e.g., \opencite{gwi}, \opencite{min}). Secondly, absorption lines are detected with the 100-m telescope that correspond to transitions which are considerably overcooled only under conditions where the masing transitions are strongly inverted (\opencite{men86}). One  such line was observed with the VLA (\opencite{wil}). So, modern interferometry in some cases can provide the basis for combined analysis of data in both maser and \lq quasi-thermal\rq\ types of lines.

In this paper we analyse the data on methanol lines obtained with the BIMA interferometer. In the first section we show that the masing region can be filled with numerous optically elongated structures which are responsible for producing maser spikes and accompanying pedestal profiles. Analysis of the data provides estimates of the physical parameters of the masing gas. In the second section we show that combined analysis of the data on the maser and \lq quasi-thermal\rq\ methanol lines reveals the physical structure of the vicinity of the W3(OH) ultracompact HII region. It is also shown that the size of the masing region exceeds that of maser hot spot sizes by orders of magnitude. In the third section we provide evidence that analysis of the \lq quasi-thermal\rq\ line data may indicate the presence of previously unknown masing objects to the south-west of W3(OH), which are to be studied by interferometrical means. In turn, the data on the H$_2$O masers around the TW object to the east of W3(OH) show the presence of molecular material which can be detected in \lq quasi-thermal\rq\ line emission.

\section{Analysis of the methanol line emission and absorption at
W3(OH)}

The BIMA interferometer was used to observe 24 methanol lines in
3mm and 1 mm wavelength ranges toward W3(OH) with a spectral
resolution better than 1.2 km/s. Details of the observations are
described in (\opencite{sut04}). This source is a prototypical
class II methanol maser source. Emission in 9 transitions which
were predicted to mase under certain conditions (\opencite{sob97})
was detected (\opencite{sut01}). Some lines display pronounced
narrow maser spike and pedestal profiles while others show the
pedestal component alone. Analysis of this data on newly
discovered masers and maser candidate lines has brought
considerable improvement to the model of the masing region in
front of W3(OH).

It was shown that maser action takes place in hot (about 150 K),
dense ($2 \times 10^{6}~\rm cm^{-3}$ ), methanol-rich gas in front of
the UCHII region, pumped by infrared radiation from warm dust.
The observed variety of maser line profiles and flux ratios
can be explained if the maser spike emission from W3(OH) arises
in the region where
maser amplification is moderate and directional, whereas for
the pedestal emission the maser action is weaker and directed
less strongly towards the observer.
This can be a result of the geometry of the maser region containing
some regions greatly elongated along the line of sight and others
elongated in different directions.
The model predicted strong absorption in 6 methanol lines with
frequencies 84.52121, 85.56807, 94.54181, 95.16952, 105.06376,
and 109.15321 GHz.

The six lines which were predicted to show absorption
in the masing region show double peaked profiles.
These are not consistent with two emission components at fixed velocities, but, rather, can be well fitted with a superposition of 2 gaussians corresponding to common emitting and absorbing regions (see figure 1). Excitation analysis in the way described in
(\opencite{sut01}) has shown that the depths of the absorptions
are well in accord with the model predictions. Figure 2 shows that
emission in the maps of absorbing lines at the maser velocities
comes from a different spatial position than does the bulk of
emission in maser lines. That means that the size of the masing
region is comparable to the total spread of the 6.7 GHz class II
methanol maser spots. Modelling of the methanol lines in the way
described in (\opencite{sut01} and \opencite{sut04}) is consistent
with the hypothesis that 2 methanol-rich regions are situated in front
of the ultracompact HII region (UCHII) in W3(OH): the first being
a rather hot (greater than 200 K) and dense (greater than
$10^{8}~\rm cm^{-3}$) methanol emitting region with angular size
slightly exceeding that of the UCHII, and the second region
producing the strong masers and absorptions. The latter has a size
of order 1 arcsecond, which is about 3 orders of magnitude greater
than that of the 6.7 GHz maser spots, and corresponds to the total
extent of the region where the strong class II methanol masers are
distributed. So, BIMA observations have shown that the methanol
maser region in W3(OH) has a size of a few by $10^{16}$ cm, and
provided estimates of the physical parameters on the basis of
combined analysis of the masing and quasi-thermally excited lines.

\begin{figure}
\centerline{\includegraphics[width=1.0\linewidth]{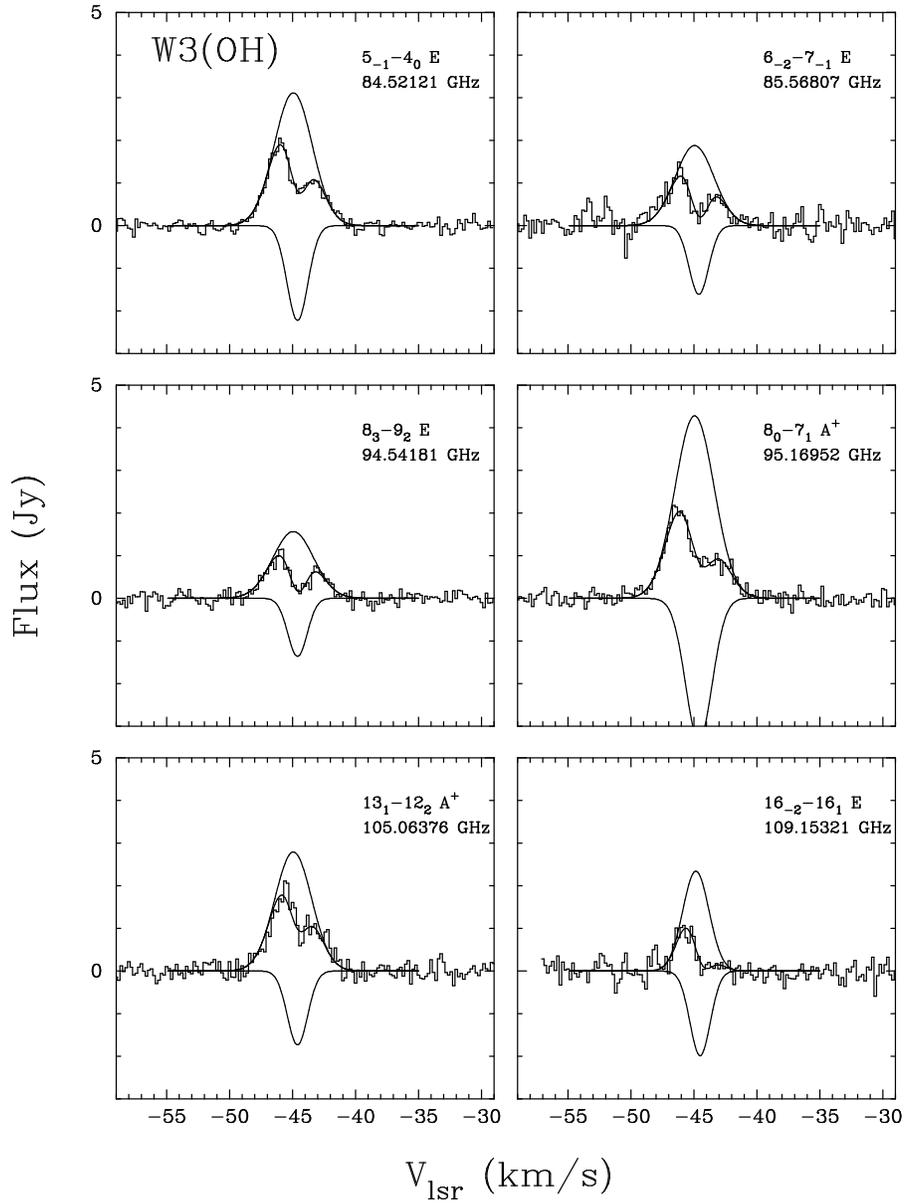}}
\caption{BIMA spectra of 6 methanol lines which were predicted to
absorb in the methanol masing region. Fits with 2 gaussians
corresponding to common emitting ($V_{LSR}=-44.9$~ km/s) and
absorbing ($V_{LSR}=-44.6$~km/s) regions are shown. 
The line profiles are not consistent with two emission components
at fixed velocities.  This suggests that there is an
absorbing region situated closer to the observer. So, it is
likely that we see absorption against the line emitting
background.} \label{w3ohspectra}
\end{figure}

\begin{figure}
\centerline{\includegraphics[width=1.0\linewidth]{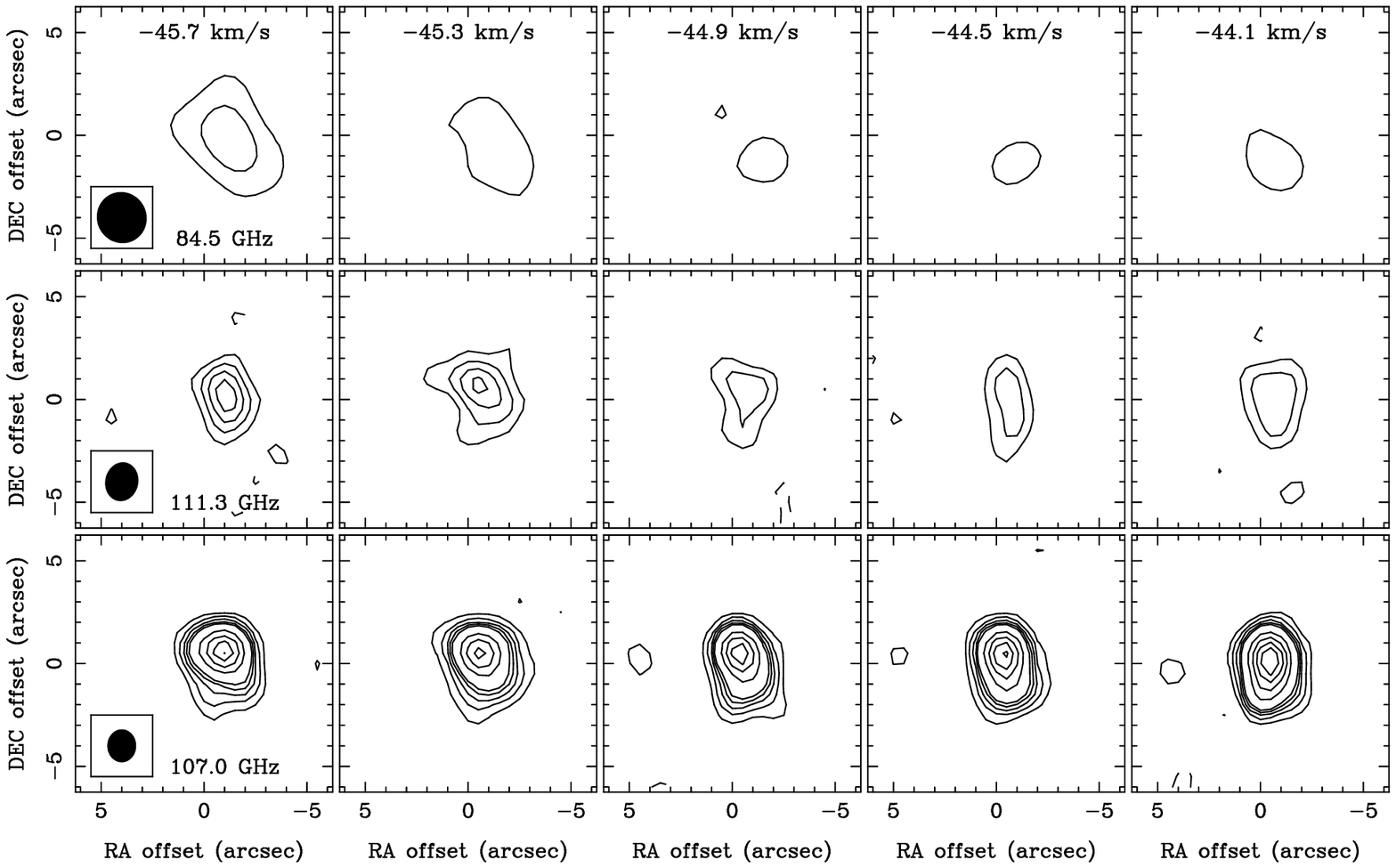}}
\caption{BIMA maps of W3(OH) in the methanol lines. The upper row of
channel maps corresponds to the line at 84.5 GHz which shows
strong absorption in the model (Sutton et al., 2001) and displays
an absorption dip in the spectrum shown in figure 1. It is seen that
no emission comes from the northern part of the source in the
central and right panels corresponding to velocities of the dip.
The middle row shows maps of the 111.3 GHz line which is transparent
in the model and shows an uneven pedestal profile. It is seen that
emission comes from both northern and southern parts of the source
at all line velocities. The lower row of panels shows maps of the
107.0 GHz line which is observed as a bright maser and behaves
accordingly in the model. It is seen that emission in the central
and right panels mostly comes from the northern part of the source
where the methanol masing region described in the text is
situated. Contour levels are 0.4, 0.8, 1.2, 1.6, 2, 4, 6, 8, and
10 Jy/beam. Interferometer beams are shown in the lower left
corners of the leftmost panels.} \label{w3ohmap}
\end{figure}

\section{How \lq quasi-thermal\rq\ line observations help reveal masers and vice versa}
Sutton et al. (2004) showed that BIMA observations reveal the presence of extended features to the south-west of W3(OH) which are traced by the methanol line emission (see also lower panels in fig.2). Methanol is a chemical tracer of shocked regions, and analysis of the spatial structure of its emission in regions of massive star formation can lead to the discovery of extremely young stellar objects displaying maser emission (\opencite{sob82}). Indeed, Sutton et al. (2004) found 2 objects which are most pronounced in the Class I methanol maser lines. Excitation analysis carried out in this paper shows that these lines are most probably weak masers, which can be confirmed by interferometric measurements with higher spatial resolution. This can be done with the forthcoming CARMA facility or with the VLA in counterpart lines of the same methanol line series.

In turn, data on the maser emission of water around the TW young stellar object situated about 6$"$ to the east of W3(OH) shows the presence of molecular material situated on both sides of the TW object with an almost equal angular separation of about 1$"$. The nature of the material in the immediate vicinity of the TW object is unclear and can be explained in terms of the jet and edge-on disk (\opencite{shc}). Water masers are produced under very special conditions and can display great deviations from the systemic velocity of the source, i.e., they might not reveal bulk motions of matter. So, observations of \lq quasi-thermal\rq\ molecular emission in the TW vicinity with appropriate angular resolution are very important to clarify the situation and, in any case, help promote understanding of important processes related to the earliest stages of massive star formation. The existence of molecular material situated 1$"$ to the west of TW was clearly shown by interferometry of red-shifted highly excited \lq quasi-thermal\rq\ line emission (\opencite{wyr97}) and a compact continuum source was found at this position (\opencite{wyr99}). However, highly excited lines and the continuum did not indicate the presence of molecular material to the east of the TW object where the strongest water masers reside.

In the paper by Sutton et al. (2004) it is shown that the TW
object is surrounded by a dense envelope of molecular gas. This
surrounding gas obscures the source interiors. However, Doppler
shifted internal objects can display themselves in the line wings.
We have searched for such emission in the wings of the strong low
excitation lines of the methanol line quartet at 96.7 GHz. Because
the lines are blended, we examined the blue wing of the 96.74458
GHz line and the red wing of the 96.73939 GHz line. Results of the
search are shown in figure 3. We found the source to the west of
TW in the red wing emission and found evidence of the existence of the
blue-shifted methanol emission from the water maser site situated
about 1$"$ east of TW. The later fact is also interesting from the
chemical point of view because it shows that the water masers are
formed in the methanol abundant region. So, masers have prompted
the detection of \lq quasi-thermal\rq\ line emission which can
provide a clue for elucidating the nature of one of the most
intriguing objects of early massive star formation.

\begin{figure}
\centerline{\includegraphics[width=1.0\linewidth]{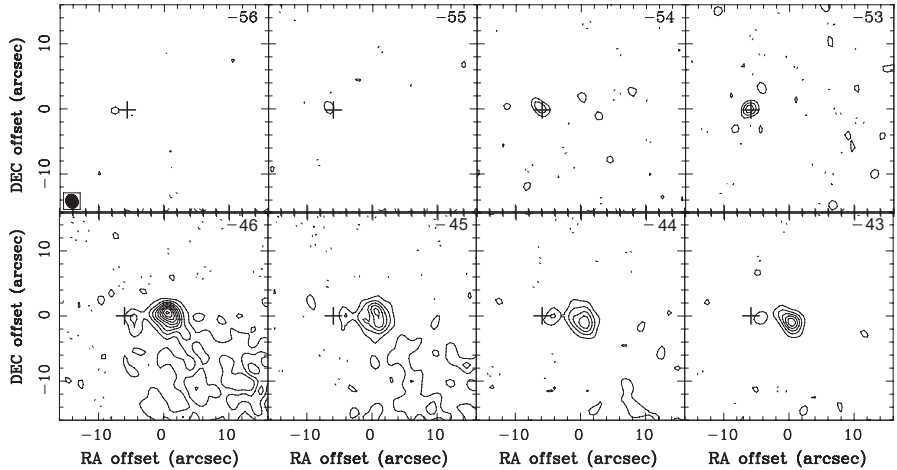}}
\caption{BIMA channel maps of the TW vicinity in the methanol
lines at 96.7 GHz. The cross marks the position of the TW object. The
upper row of panels shows emission in the blue wing of the
96.74458 GHz line and the leftmost upper panel displays embedded
material situated about 1$"$ east of the TW. The lower row of
panels corresponds to the red wing of the 96.73939 GHz line and
displays the embedded source to the west of the TW.  The interferometer
beam is shown in the lower left corner of the leftmost upper
panel.} \label{TWmap}
\end{figure}

\section{Conclusions}
This paper provides an example of how combined analysis of the
interferometry data on the maser and \lq quasi-thermal\rq\ lines helps
to elucidate the structural, physical and chemical status of the
regions of massive star formation. Data on sets of both maser and
\lq quasi-thermal\rq\ lines allows estimates of the temperature,
density, molecular abundances and source extensions and shapes. We
find that the numerous methanol transitions form a good basis for
such types of studies both from the point of view of their high
diagnostic capacity and because relevant lines are rather easy to
observe with existing and forthcoming radioastronomical
facilities. We have also shown that information on \lq quasi-thermal\rq\
lines can reveal the existence of masing objects while the masers can
prompt the detection of molecular material emitting in \lq quasi-thermal\rq\
lines.

\begin{acknowledgements}
AMS was supported by grants from RFBR (03-02-16433) and the Russian
Ministry of Education (E02-11.0-43). ECS was supported by NSF
Grant AST 02-28953. DMC and PDG were supported by the Australian Research Council.
\end{acknowledgements}

\end{article}
\end{document}